\documentclass{epl}
\usepackage{bm,times,amsmath,amssymb,amsfonts,pstricks}
\usepackage{times}
\newcommand{\openone}{\leavevmode\hbox{\small1\normalsize\kern-.33em1}}
\title{Fluctuation-induced forces in periodic slabs: Breakdown of
  \mth{\epsilon} expansion at the bulk critical point and revised field
  theory}
\shorttitle{Fluctuation-induced forces in periodic slabs}
\author{H.~W.\ Diehl\inst{1} \and Daniel Gr{\"u}neberg\inst{1} \and
  M.~A. Shpot\inst{1,2}}
\institute{%
  \inst{1} Fachbereich Physik, Universit{\"a}t Duisburg-Essen, Campus
  Essen, D-45117 Essen, Germany\\
\inst{2} Institute for Condensed Matter Physics, 79011 Lviv, Ukraine}

\shortauthor{H.~W.\ Diehl \etal}
\pacs{05.70.Jk}{Critical point phenomena} 
\pacs{68.15.+e}{Liquid thin films}
\pacs{11.10.-z}{Field theory}

\begin{document}

\maketitle

\begin{abstract}
 Systems described by $n$-component $\phi^4$ models in a
  $\infty^{d-1}\times L$ slab geometry of finite thickness $L$ are
  considered at and above their bulk critical temperature
  $T_{c,\infty}$. The renormalization-group improved perturbation
  theory commonly employed to investigate the fluctuation-induced
  forces (``thermodynamic Casimir effect'') in $d=4-\epsilon$ bulk
  dimensions is re-examined. It is found to be ill-defined beyond
  two-loop order because of infrared singularities when the boundary
  conditions are such that the free propagator in slab geometry
  involves a zero-energy mode at bulk criticality. This applies to
  periodic boundary conditions and the special-special ones corresponding
  to the critical enhancement of the surface interactions on both
  confining plates. The field theory is reorganized such that a
  small-$\epsilon$ expansion results which remains well behaved down
  to $T_{c,\infty}$. The leading contributions to the critical Casimir
  amplitudes $\Delta_{\mathrm{per}}$ and $\Delta_{\mathrm{sp},\mathrm{sp}}$
  beyond two-loop order are $\sim (u^*)^{(3-\epsilon)/2}$, where
  $u^*=O(\epsilon)$ is the value of the renormalized $\phi^4$ coupling
  at the infrared-stable fixed point. Besides integer powers of
  $\epsilon$, the small-$\epsilon$ expansions of these amplitudes
  involve fractional powers $\epsilon^{k/2}$, with $k\geq 3$, and
  powers of $\ln \epsilon$. Explicit results to order
  $\epsilon^{3/2}$ are presented for $\Delta_{\mathrm{per}}$ and
  $\Delta_{\mathrm{sp},\mathrm{sp}}$, which are used to estimate their
  values at $d=3$.
\end{abstract}

Fluctuations associated with long wave-length, low-energy excitations
play a crucial role in determining the physical properties of many
macroscopic systems. When such fluctuations are confined by
boundaries, walls, or size restrictions along one or several
axes, important effective forces may result. In those cases where the
continuous mode spectrum that emerges as the system becomes
macroscopic in all directions is not separated from zero energy by a
gap, these fluctuation-induced forces are long-ranged, decaying
algebraically as a function of the relevant confinement length $L$
(separation of walls, thickness of the system, etc).

A prominent example of such forces are the Casimir forces
\cite{Cas48} induced by vacuum fluctuations of the electromagnetic
field between two metallic bodies a distance $L$ apart
\cite{BMM01,Lam97,MR98}.  Analogous long-range effective forces occur
in condensed matter systems as the result of either (i) thermal
fluctuations at continuous phase transitions or else (ii) Goldstone
modes and similar ``massless'' excitations
\cite{FdG78,Sym81,NI85,KD91,KD92a,Kre94,KG99,BDT00}. In particular the
former ones, frequently called ``critical Casimir forces,''
have attracted considerable theoretical and experimental attention
recently. Beginning with the seminal paper by Fisher and de Gennes
\cite{FdG78}, they have been studied theoretically for more than a
decade using renormalization group (RG)
\cite{Sym81,NI85,KD91,KD92a,Kre94} and conformal field theory methods
\cite{DiFMS97}, exact solutions of models \cite{BDT00}, as
well as Monte Carlo (MC) simulations \cite{KL96,DKD03,DK04}.  Though their
detailed experimental investigation just began, recent experiments
\cite{GC99,GC02,UCH01,IB05,Fyp05} have provided clear evidence of
their occurrence.

The goal of the present work is to reexamine the existence of the
$\epsilon=4-d$ expansion for critical Casimir forces. Considering
$n$-component $\phi^4$ models in $\infty^{d-1}\times L$ slab
geometries, subject to different kinds of large-scale boundary
conditions (BC), we shall show that for some of these BC, the
$\epsilon$ expansion breaks down at the bulk critical temperature
$T_{c,\infty}$ beyond two-loop order because of infrared
singularities. This breakdown occurs quite generally whenever the
two-point correlation function in Landau theory has a zero mode at
bulk criticality, as it does for periodic BC and for those
corresponding to critical enhancements of the short-range surface
interaction at both walls \cite{Die86a,Die97}. 

As we shall show, the infrared problems one has in these cases require
a special treatment of the zero mode.  This leads to a modified
RG-improved perturbation expansion, which is well-defined at and above
$T_{c,\infty}$, but yields contributions to the critical Casimir force
of the form $(u^*)^{(3-\epsilon)/2}$ beyond two-loop order, where
$u^*=O(\epsilon)$ is the value of the renormalized $\phi^4$ coupling
constant at the infrared-stable fixed point in $d=4-\epsilon$ bulk
dimensions. Thus, contrary to common belief, this quantity does not
generally have an expansion in integer powers of $\epsilon$; the
present theory requires fundamental revision.

That zero modes may play an important role and require special
treatment has been observed before in studies of finite-size effects
in systems that are finite in all, or in all but one, directions
\cite{zeromodes}. Such systems (with short-range interactions) differ
from the ones in slab geometry considered here in that they do not
have a sharp phase transition at $T>0$, except in the bulk limit.

To put things in perspective, consider a system in a
$\infty^{d-1}\times L$ slab geometry whose bulk (${L\to\infty}$)
critical behavior is representative of the universality class of the
$d$-dimensional $n$-component $\phi^4$ model with short-range
interactions. Its reduced free energy per unit cross-sectional area
$A\to \infty$ can be decomposed as
\begin{equation}
  \label{eq:fdec}
  f_L\equiv\lim_{A\to\infty}\frac{F}{Ak_BT_{c,\infty}}
  =L\,f_{\text{bk}}+f_{s,a}+f_{s,b}+f^{\text{res}}_{a,b}(L)\;,
\end{equation}
where $f_{\text{bk}}$ is the reduced bulk free energy per volume,
while $f_{s,a}+f_{s,b}$ is the surface excess free energy that results
as the separation $L$ of the two confining plates is increased to
infinity. According to the theory of finite-size scaling, the residual
free energy  should take the
scaling form
\begin{equation}
  \label{eq:fressf}
  f^{\text{res}}_{a,b}(L)\approx
  L^{-(d-1)}\,\Theta_{a,b}(L/\xi_\infty)
\end{equation}
on sufficiently large length scales, where $\xi_\infty$ is the bulk
correlation length. Here $\Theta_{a,b}$ is a universal scaling
function, which depends on the bulk universality class, the geometry
of the system, and gross properties of the confining planes, such as
large-scale boundary conditions (BC) associated with RG fixed points of the
corresponding boundary field theory, but is independent of microscopic
details. Its value at bulk criticality,
$\Delta_{a,b}\equiv\Theta_{a,b}(0)$, is a universal number, the
so-called Casimir amplitude.

The so far most comprehensive and detailed theoretical investigation
of critical Casimir forces is that of Krech and Dietrich (KD)
\cite{KD91,KD92a} who considered $n$-component $\phi^4$ models in a
$\infty^{d-1}\times L$ slab geometry whose Hamiltonian is given by
\begin{equation}
  \label{eq:ham}
 \mathcal{H}[\phi]=\int_0^L
 dz\int_{\mathbb{R}^{d-1}}{\!\!}d^{d-1}r\left[\frac{1}{2} (\nabla\phi)^2
   +\frac{\mathring{\tau}}{2}\,\phi^2+\frac{\mathring{u}}{4!}\,\phi^4\right]
\end{equation}
together with BC. KD considered periodic ($\mathrm{per}$) and antiperiodic
($\mathrm{ap}$) BC, as well as $(a,b)=(D,D)$, $(D,\mathrm{sp})$, and
$(\mathrm{sp},\mathrm{sp})$.  The notations $D$ and $\mathrm{sp}$
indicate Dirichlet and special BC on the boundary planes
$\mathfrak{B}_1$ at ${z=0}$ and $\mathfrak{B}_2$ at $z=L$. The
Dirichlet BC corresponds to the infrared-stable fixed point describing
the surface critical behavior at the ordinary transition of
semi-infinite systems.  By special, the large-scale BC on a plate is
meant that applies when the surface interactions on it are critically
enhanced and no symmetry breaking surface terms are present there;
they pertain to the fixed point associated with the so-called special
transition \cite{Die86a}. All combinations
$(a,b)$ of such BC with $a,\,b=D,\,\mathrm{sp}$ can be
implemented by adding surface terms
$\sum_{j=1}^2\int_{\mathfrak{B}_j}\mathring{c}_j\,\phi^2/2$ to the
Hamiltonian (\ref{eq:ham}) with $\mathring{c}_j=\infty$ or
$\mathring{c}_j=\mathring{c}_{\mathrm{sp}}$ (critical enhancement
\cite{rem:ce}), depending on whether $a,\,b=D,\,\mathrm{sp}$.

Performing two-loop calculations of the excess free energies, Casimir
forces, and their scaling functions for the BC $\wp=\mathrm{per}$,
$\mathrm{ap}$, $(D,D)$, $(D,\mathrm{sp})$, and
$(\mathrm{sp},\mathrm{sp})$, and restricting themselves to
temperatures $T\geq T_{c,\infty}$, KD obtained expansions to first
order in $\epsilon$ of the Casimir amplitudes $\Delta_{\wp}$ and the
scaling functions $\Theta_{\wp}(y)$.

The restriction $T\geq T_{c,\infty}$ had not only technical reasons.
As $T$ is lowered beneath $T_{c,\infty}$, a crossover to the critical
behavior of an effectively $d-1$~dimensional system is expected to
occur at a temperature $T_{c,L}<T_{c,\infty}$. The
$\epsilon$~expansion is unable to deal with the $d-1$~dimensional
infrared singularities at $T_{c,L}$ since the appropriate small
dimensional parameter would be $\epsilon_5\equiv 5-d$. KD were aware
of this problem. They verified that their $O(\epsilon)$~results for
the boundary conditions $\wp=\mathrm{ap},\,(D,D)$, and
$(D,\mathrm{sp})$ were consistent with the asymptotic behavior of the
scaling functions $\Theta_\wp(y)$ for $y\to 0\pm$ one can infer from the
requirement that $f_L$ be analytic at $T_{c,\infty}$ when $L<\infty$.
They also noted that one could not expect their results for
$\Theta_{\mathrm{per}}$ and $\Theta_{\mathrm{sp},\mathrm{sp}}$ to
exhibit such a small-$y$ behavior because the corresponding free
propagators with periodic and Neumann boundary conditions could not be
analytically continued to negative values of $\tau\sim
(T-T_{c,\infty})/T_{c,\infty}$, unlike those for the other boundary
conditions. In order to assess the predictive power of their
$O(\epsilon)$ results for $\Theta_\wp$, KD (and others
\cite{BDT00,DKD05}) nevertheless extrapolated these functions for all
five types of boundary conditions to $d=3$ by setting $\epsilon=1$.

To see that the $\epsilon$~expansion breaks down at $T=T_{c,\infty}$
for $\wp=\mathrm{per}$ and $(\mathrm{sp},\mathrm{sp})$, consider the
three-loop graph
\unitlength=0.1em
\begin{picture}(30,10)(-5,-3)\thicklines
\put(0,0){\circle{10}} \put(5,0){\circle*{3.5}} \put(15,0){\circle*{3.5}}
\put(10,0){\circle{10}}
\put(20,0){\circle{10}}
\end{picture}
%
% and
% \psset{unit=0.08em}
% \begin{pspicture}(-17,-4)(18,8)
%  \psellipse[linewidth=0.7pt](0,-1)(16,8)
%  \psellipse[linewidth=0.7pt](0,-1)(16,4)
% \pscircle*(-15,-1){2}
% \pscircle*(15,-1){2}
% \end{pspicture}
of $f_L$, where the lines represent the free propagator
\begin{equation}
  \label{eq:GL}
  G^{(L)}_\wp(\bm{x};\bm{x}')=\int \frac{d^{d-1}p}{(2\pi)^{d-1}}
  \sum_m\frac{\langle z|m\rangle\langle m|
    z'\rangle}{p^2+k_m^2+\mathring{\tau}}\,e^{i\bm{p}\cdot(\bm{r}-\bm{r}')}
\end{equation}
between the points $\bm{x}=(\bm{r},z)$ and $\bm{x}'=(\bm{r}',z')$.
Here $\langle z|m\rangle$ and $k_m^2$ are the orthonormal
eigenfunctions and eigenvalues for the BC $\wp$, respectively. For
example, for $\wp=\mathrm{per}$ one has $\langle
z|m\rangle=L^{-1/2}\,e^{ik_mz}$ with $k_m=2\pi m/L$ and $m\in
\mathbb{Z}$, whereas $\langle
z|m\rangle=\sqrt{(2-\delta_{m,0})/L}\,\cos(k_mz)$ for
$\wp=(\mathrm{sp},\mathrm{sp})$, with ${k_m=\pi m/L}$, $m=0,1,
\ldots,\infty$.

The central part of this three-loop graph is the subgraph
\unitlength=0.1em
\begin{picture}(13,10)(3,-3)\thicklines
\put(10,0){\circle{10}} \put(5,0){\circle*{3.5}}
\put(15,0){\circle*{3.5}} \put(5,0){\line(-2,1){4}}
\put(5,0){\line(-2,-1){4}}  \put(15,0){\line(2,1){4}}
\put(15,0){\line(2,-1){4}}
\end{picture}\,,
which behaves in the bulk case $L=\infty$ (for zero external momenta) as
$\mathring{\tau}^{-\epsilon/2}$, and hence is \emph{infrared singular} at
$T_{c,\infty}$ \cite{rem:tc}. However, each of the two other bubbles
contributes a factor $G^{(\infty)}(\bm{x};\bm{x})\sim
\mathring{\tau}^{1-\epsilon/2}$. Hence, the limit $\mathring{\tau}\to
0$ of the overall contribution vanishes \cite{rem:tc}. When
$L<\infty$, each of the two tadpoles $G^{(L)}_{\wp}(\bm{x};\bm{x})$
for $\wp=\mathrm{per},\,(\mathrm{sp},\mathrm{sp})$ can be decomposed
into a contribution $(P_0G^{(L)}_{\wp}P_0)(\bm{x};\bm{x})\sim
\mathring{\tau}^{(1-\epsilon)/2}$ from the ${k_m=0}$~mode and a
remainder $(Q_0G^{(L)}_{\wp}Q_0)(\bm{x};\bm{x})$, where
$P_0=\openone-Q_0=|0\rangle\langle 0|$. At $\mathring{\tau}=0$ we are
left with the contributions from the remainder, namely
\begin{equation}
  \label{eq:Gxxper}
  (Q_0G^{(L)}_{\mathrm{per}}Q_0)(\bm{x};\bm{x})|_{\mathring{\tau}=0}=
    \frac{\Gamma(1-\epsilon/2)}{2\pi^{2-\epsilon/2}}\,
  \frac{\zeta(2-\epsilon)}{L^{2-\epsilon}}
\end{equation}
and
\begin{eqnarray}
  \label{eq:Gxxsp}
(Q_0G^{(L)}_{\mathrm{sp},\mathrm{sp}}Q_0)(\bm{x};\bm{x})|_{\mathring{\tau}=0}=
\frac{\Gamma(1-\epsilon/2)}{2^{4-\epsilon}\pi^{2-\epsilon/2}\,L^{2-\epsilon}}\,
\Big[2\,\zeta(2-\epsilon)+\zeta\Big(2-\epsilon,\frac{z}{L}\Big) +\zeta\Big(2-\epsilon,\frac{L-z}{L}\Big)\Big],
\end{eqnarray}
where $\zeta(s,a)=\sum_{j=0}^\infty(j+a)^{-s}$ is the Hurwitz zeta
function.

The zero-mode contribution $[(P_0G^{(L)}_\wp P_0)(\bm{x};\bm{x}')]^2$ of
\begin{picture}(13,10)(3,-3)\thicklines
\put(10,0){\circle{10}} \put(5,0){\circle*{3.5}}
\put(15,0){\circle*{3.5}} \put(5,0){\line(-2,1){4}}
\put(5,0){\line(-2,-1){4}}  \put(15,0){\line(2,1){4}}
\put(15,0){\line(2,-1){4}}
\end{picture}
must be integrated over the parallel separation $\bm{r}-\bm{r}'$. The
result behaves $\sim\mathring{\tau}^{-(1+\epsilon)/2}$. Combined with
the $(Q_0G^{(L)}_{\wp}Q_0)$ parts of the tadpoles, it produces a
contribution that diverges as $\mathring{\tau}\to 0$. Thus the loop
expansion is ill-defined at $\mathring{\tau}=0$ for these two BC
$\wp=\mathrm{per}$ and $(\mathrm{sp},\mathrm{sp})$. Evidently, this
breakdown should occur more generally whenever a $k_m=0$~mode is
present, which is not the case for $\wp=\mathrm{ap},\,(D,D)$, and
$(D,\mathrm{sp})$.

The origin of this breakdown is a deficiency of Landau theory:
Whenever the free propagator involves a $k_m=0$~mode, it predicts a
transition for the bulk and the film of finite thickness $L$ at the
same critical value $\mathring{\tau}=0$. The remedy is a reorganization of the
perturbation series. The ${k_m=0}$~mode must be split off and treated in
the background of the $k\neq 0$~modes. As a result of its coupling to
the latter, the zero mode becomes massive at $T_{c,\infty}$ for
$L<\infty$.

To formulate such an expansion, we decompose $\phi$ as
$\phi(\bm{x})=\varphi(\bm{r})+\psi(\bm{x})$ into its $k=0$~mode
contribution $\varphi(\bm{r})$ and its orthogonal component $\psi$
with $\int_0^L dz\,\psi(\bm{r},z)=0$. Integrating out $\psi$ defines
us an effective $d-1$~dimensional field theory with the Hamiltonian
\begin{equation}\label{eq:heff}
  \mathcal{H}_{\mathrm{eff}}[\varphi]=-\ln\text{Tr}_\psi
  e^{-\mathcal{H}[\varphi+\psi]}\,. 
\end{equation}
Let us introduce the free-energy part $F_\psi$ and the average
$\langle\ldots \rangle_\psi$ by
\begin{equation}
  \label{eq:fep}
  F_\psi\equiv -\ln \text{Tr}_\psi e^{-\mathcal{H}[\psi]}\;,\qquad
  \langle\ldots \rangle_\psi\equiv e^{F_\psi}\, 
\text{Tr}_\psi\Big(\ldots e^{-\mathcal{H}[\psi]}\Big)\,.
\end{equation}
Then we have
\begin{equation}
  \label{eq:Heff}
  \mathcal{H}_{\mathrm{eff}}[\varphi]=F_\psi+\mathcal{H}[\varphi]-\ln\langle
  e^{-\mathcal{H}_{\text{int}}[\varphi,\psi]} \rangle_\psi
\end{equation}
with
\begin{equation}
  \label{eq:Hint}
  \mathcal{H}_{\text{int}}[\varphi,\psi]\equiv \int_0^L
  dz\int_{\mathbb{R}^{d-1}}
  d^{d-1}r\,\Big[\frac{\mathring{u}}{4}\,\varphi^2\psi^2
    +\frac{\mathring{u}}{6}\,(\bm{\varphi}\cdot\,\bm{\psi})\,\psi^2\Big]
\end{equation}
and
\begin{equation}
  \label{eq:heff0}
  \mathcal{H}[\varphi]
  =\int_{\mathbb{R}^{d-1}}d^{d-1}r\,
  \Big[\frac{L}{2}(\partial_{\bm{r}}\varphi)^2
    +\frac{\mathring{\tau} L}{2}\,\varphi^2+\frac{\mathring{u} L}{4!}\,\varphi^4\Big].
\end{equation}
The last term in Eq.~(\ref{eq:Heff}) gives loop corrections
$\sum_{l=1}^\infty \mathcal{H}_{\mathrm{eff}}^{[l]}[\varphi]$. For the one-loop
contribution, one obtains
\begin{eqnarray}
  \label{eq:heff1}
\mathcal{H}_{\mathrm{eff}}^{[1]}[\varphi]&=&\frac{1}{2}\,\text{Tr}
 \ln\big[\openone+(\mathring{u}/6)\,G^{(L)}_\psi
   \big(\delta_{\alpha\beta}\,\varphi^2
     +2\varphi_\alpha\varphi_\beta\big)\big]\nonumber\\
&=& -
\psset{unit=2.2em}
\begin{pspicture}[0](-1,0)(1,0.7)
\psline[linewidth=1pt,linecolor=gray]{-}(-0.4,-0.0)(0.4,-0.0)
\pscircle[linewidth=0.7pt,linestyle=dashed](0,0.38){0.4}
\rput(-0.6,0){$\varphi$} \rput(0.6,0){$\varphi$}
 \pscircle*(0,0){0.1}
\end{pspicture}-
\raisebox{2pt}{\begin{pspicture}[0](-1.1,0)(1.8,0.7)
\psline[linewidth=1pt,linecolor=gray]{-}(-0.4,0)(-0.6,0.2)
\psline[linewidth=0.7pt,linecolor=gray]{-}(-0.4,0)(-0.6,-0.2)
\psellipse[linewidth=0.7pt,linestyle=dashed](0.38,0)(0.8,0.4)
\psline[linewidth=1pt,linecolor=gray]{-}(1.16,0.)(1.36,0.2)
\psline[linewidth=1pt,linecolor=gray]{-}(1.16,0)(1.36,-0.2)
\rput(-0.8,0.3){$\varphi$} \rput(-0.8,-0.3){$\varphi$}
\rput(1.6,0.3){$\varphi$} \rput(1.6,-0.3){$\varphi$}
\pscircle*(-0.4,0){0.1} \pscircle*(1.16,0){0.1}
\end{pspicture} }+\ldots,\qquad
\end{eqnarray}
where the dashed lines represent free $\psi$-propagators
$G^{(L)}_\psi=Q_0G^{(L)}_\wp Q_0$, while the gray bars indicate
$\varphi$~legs.

Added to $\mathcal{H}[\varphi]$, the first graph in Eq.~(\ref{eq:heff1})
produces the shift
\begin{equation}
  \label{eq:tbshift}
  \mathring{\tau}\to\mathring{\tau}^{(L)}_\wp\equiv
\mathring{\tau}+\delta\mathring{\tau}^{(L)}_\wp\quad\text{with}\quad
\delta\mathring{\tau}^{(L)}_\wp=\frac{\mathring{u}}{2}\,
\frac{n+2}{3}\,\int_0^L\frac{dz}{L}\, 
  (Q_0G_\wp^{(L)}Q_0)(\bm{x};\bm{x}) \;,
\end{equation}
so that the free $\varphi$-propagator $G^{(L)}_\varphi$ acquires an
$L$-dependent mass at ${\mathring{\tau}=0} $. The second graph in
Eq.~(\ref{eq:heff1}) yields a nonlocal $\varphi^2\varphi^2$
interaction; the suppressed ones correspond to similar nonlocal
interactions involving more than two $\varphi^2$ operators. The
two-loop term $\mathcal{H}_{\mathrm{eff}}^{[2]}[\varphi]$ involves two contributions ${\sim
\varphi\varphi}$; a local one giving an $O(\mathring{u}^2)$ correction
to the shift $\delta\mathring{\tau}^{(L)}_\wp$, and a nonlocal one whose
interaction potential is proportional to
$\int_0^L dz\,\int_0^L dz'\,[G_\psi(\bm{x},\bm{x}')]^3$.

Upon substituting the results (\ref{eq:Gxxper}) and (\ref{eq:Gxxsp})
into Eq.~(\ref{eq:tbshift}), the shifts
$\delta\mathring{\tau}^{(L)}_{\mathrm{per}}$ and
$\delta\mathring{\tau}^{(L)}_{\mathrm{sp},\mathrm{sp}}$ can be
computed in a straightforward manner. At $\mathring{\tau}=0$, one
obtains
\begin{equation}
  \label{eq:shiftper}
   \delta\mathring{\tau}^{(L)}_{\mathrm{per}}= 2^{2-\epsilon}\,
   \delta\mathring{\tau}^{(L)}_{\mathrm{sp},\mathrm{sp}}
= \mathring{u}\,\frac{n+2}{6}\,
\frac{\Gamma(1-\epsilon/2)\, \zeta(2-\epsilon)}{2\pi^{2-\epsilon/2}\,
  L^{2-\epsilon}}.
\end{equation}

We can now set up a Feynman graph expansion, utilizing
$\hat{G}_\varphi(\bm{p})=\big[L\big(p^2
+\mathring{\tau}_\wp^{(L)}\big)\big]^{-1}$ as free propagator in the
${d-1}$~dimensional momentum space and employing dimensional
regularization. Since $\hat{G}_\varphi(\bm{p})$ remains massive at
$\mathring{\tau}=0$ when $L<\infty$, the Feynman graphs are infrared
finite as long as $\mathring{\tau}\geq 0$. Owing to our reorganization
of perturbation theory, the breakdown encountered in the conventional
expansion in terms of $G_\wp^{(L)}$ is avoided. Note also that the
correct bulk expressions are recovered as $L\to \infty$ since the
contributions from the zero mode vanish. For example, the term
$F_\psi$ in Eq.~(\ref{eq:Heff}) yields the bulk free energy when
$L\to\infty$. Indeed, setting $T=T_{c,\infty}$ and denoting the analog
of $f^{\mathrm{res}}_\wp$ for $F_\psi$ as
$f_{\psi;\wp}^{\mathrm{res}}$ we find
\begin{equation}
  \label{eq:fpsires}
  \frac{L^{d-1}}{n}\,f_{\psi;\wp}^{\mathrm{res}}\big|_{T_{c,\infty}}=
  a^{(0)}_\wp(\epsilon)+\frac{n+2}{4!}\,a^{(1)}_\wp(\epsilon)\,\mathring{u}
  L^\epsilon +O(\mathring{u}^2 L^{2\epsilon})
\end{equation}
with
\begin{eqnarray}
  \label{eq:aps}
  a^{(0)}_{\mathrm{per}}(\epsilon)&=&2^{4-\epsilon}\,
  a^{(0)}_{\mathrm{sp},\mathrm{sp}}(\epsilon)=
  -\frac{\Gamma(2-\epsilon/2)}{\pi^{2-\epsilon/2}}\,\,\zeta(4-\epsilon) \,,\nonumber\\
a^{(1)}_{\mathrm{per}}(\epsilon)&=&2^{-2}\pi^{\epsilon-4}\,
\Gamma^2(1-\epsilon/2)\,\zeta^2(2-\epsilon) \,,\nonumber\\
a^{(1)}_{\mathrm{sp},\mathrm{sp}}(\epsilon)&=&\frac{\pi^{1-\epsilon}}{4^{3-\epsilon}}
\frac{2\,\zeta^2(\epsilon-1)
  +\zeta(2\epsilon-2)}{2\cos^2(\pi\epsilon/2)\,
  \Gamma^2[(3-\epsilon)/2]} \,.
\end{eqnarray}

KD's two-loop results for $f_{\wp}|_{T_{c,\infty}}$ with
$\wp=\mathrm{per},\, (\mathrm{sp},\mathrm{sp})$ follow from
Eqs.~(\ref{eq:fpsires}) and (\ref{eq:aps}), as they should because the
zero-mode contributions to $f^{\mathrm{res}}_\wp$ of order $\mathring{u}^0$ and
$\mathring{u}$ vanish at $T_{c,\infty}$.
However, the $\varphi$-dependent part of
$\mathcal{H}_{\mathrm{eff}}[\varphi]$ gives additional contributions.
The leading ones correspond to one- and two-loop terms of a
$\varphi^4$ theory in $d-1$ dimensions with mass coefficient
$\mathring{\tau}_\wp^{(L)}$ and coupling constant $\mathring{u}/L$, as
the rescaling $L^{1/2}\varphi\to\varphi$ in Eq.~(\ref{eq:heff0})
shows. At $T_{c,\infty}$, they yield $L$-dependent contributions to
$f^{\mathrm{res}}_\wp$ proportional to
$(\delta\mathring{\tau}_\wp^{(L)})^{(3-\epsilon)/2}\sim
L^{1-d}(\mathring{u} L^\epsilon)^{(3-\epsilon)/2}$ and
$(\mathring{u}/L)(\delta\mathring{\tau}_\wp^{(L)})^{1-\epsilon}\sim
L^{1-d}\,(\mathring{u} L^\epsilon)^{2-\epsilon}$, respectively.
Including only the first one, we arrive at
\begin{equation}
  \label{eq:fres}
  \frac{L^{d-1}}{n}\big[f_{\wp}^{\mathrm{res}}
  -f_{\psi;\wp}^{\mathrm{res}}\big]_{T_{c,\infty}}= A_\wp(\epsilon) \,
  \Big(\frac{n+2}{4!} \mathring{u} L^\epsilon \Big)^{(3-\epsilon)/2} +\ldots
\end{equation}
with
\begin{eqnarray}
  \label{eq:Awp}
  A_{\mathrm{per}}(\epsilon)=2^{(2-\epsilon)(3-\epsilon)/2}
  A_{\mathrm{sp},\mathrm{sp}}(\epsilon)
    & =& -\frac{\Gamma[(\epsilon-3)/2]}{2^{4-\epsilon}}
  \bigg[\frac{2\Gamma(1-\epsilon/2)\,
\zeta(2-\epsilon)}{\pi^{(6-\epsilon)/2}}\bigg]^{(3-\epsilon)/2} .\quad
\end{eqnarray}

To combine these results with RG-improved perturbation theory, we
follow KD. We utilize the reparametrizations
($\mathring{u}=2^d\pi^{d/2}Z_u\mu^\epsilon u$,
$\mathring{\tau}=\mu^2Z_\tau \tau$, \ldots) of bulk and surface
quantities of the corresponding semi-infinite theories \cite{Die86a},
and fix the additional additive (bulk and surface) counterterms $f_L$
requires such that $f_{L,\mathrm{ren}}$, its renormalized counterpart,
vanishes at ${\tau=1}$ together with its 1st and 2nd
$\tau$-derivatives \cite{rem:cderi}. To obtain the critical Casimir
amplitudes $\Delta_\wp$, we must express $f^{\mathrm{res}}_\wp$ in
terms of renormalized variables, set $\mu L=1$ and $\tau=0$, and
evaluate it at the fixed-point value $u^*=3\epsilon/(n+8) +
O(\epsilon^2)$. Upon expanding in powers of $\epsilon$, we find
($\gamma=$ Euler-Mascheroni constant)
\begin{equation}
  \label{eq:Delper}
  \frac{\Delta_{\mathrm{per}}}{n} =  -\frac{\pi^{2}}{90}
  +\frac{\pi^{2}\epsilon}{180}\bigg[1-\gamma-\ln \pi
    +\frac{2\zeta^{\prime}(4)}{\zeta(4)}
     +\frac{5}{2}\frac{n+2}{n+8}
   \bigg]
 -\frac{\pi^{2}}{9\sqrt{6}} \left
   (\frac{n+2}{n+8}\right)^{3/2}\epsilon^{3/2}+O(\epsilon^{2})
\end{equation}
and
\begin{eqnarray}\label{eq:Delspsp}
\frac{\Delta_{\mathrm{sp},\mathrm{sp}}}{n} &=& -\frac{\pi^{2}}{1440}
+\frac{\pi^{2}\epsilon}{2880}\bigg[1-\gamma
-\ln(4\pi)+\frac{5}{2}\frac{n+2}{n+8}
+\frac{2\zeta^{\prime}(4)}{\zeta(4)}\bigg] 
 \nonumber \\ && \strut
 -\frac{\pi^{2}}{72\sqrt{6}} \bigg(\frac{n+2}{n+8}
 \bigg)^{3/2}\epsilon^{3/2}+O(\epsilon^{2})\;.
\end{eqnarray}

The $O(\epsilon^{3/2})$ terms result from the $O(
\mathring{u}^{(3-\epsilon)/2})$ contributions in Eq.~(\ref{eq:fres}).
Obviously, the latter also implies contributions of the form
$\epsilon^{k+3/2}\ln^k\epsilon$ with $k\in\mathbb{N}$. Furthermore,
the terms $(\mathring{u}/L)(\delta\mathring{\tau}_\wp^{(L)})^{1-\epsilon}\sim \mathring{u}^{2-\epsilon}$
mentioned above yield contributions of the form
$\epsilon^{k+2}\ln^k\epsilon$.

Can the appearance of the $\epsilon^{3/2}$ and unconventional
higher-order terms be checked by alternative means? This is indeed
possible: The limiting value $\lim_{n\to\infty} \Delta_{\mathrm{per}}
/n$ can be obtained from the exact solution of the mean-spherical
model (expressed in terms of the function $Y_0$ of \cite{DDG06}, it
becomes $Y_0(0,0)$).  Solving the corresponding self-consistent
equations iteratively with $\epsilon>0$ reproduces the $n\to\infty$
limit of all terms on the right-hand side of Eq.~(\ref{eq:Delper})
and shows the existence of higher-order contributions of the mentioned
form \cite{DGSD06}.

In Table~\ref{tab:Deltaest} we give the values
\begin{table}[htb]
  \caption{Estimated values of $\Delta_\wp(d,n)/n$ for $d=3$.
    Following \cite{KD92a}, we have included the $d=3$ values of
    $\Delta_{\mathrm{sp},\mathrm{sp}}$, although they are
    probably of little physical interest because the special surface
    transition is not expected to occur at $d=3$ in the $O(n\geq 2)$ case,
    unless surface couplings of infinite strengths are allowed.}
\label{tab:Deltaest}
\begin{center}
\begin{largetabular}{ccccc}\hline\hline
$n$&$1$&$2$&$3$&$\infty$\\ \hline
$\Delta_{\mathrm{per}}(3,n)/n$&$-0.1967^a$&$-0.2147^a$&
$-0.2311^a$&$-0.4668^a$\\
&$-0.1105^b$&$-0.1014^b$&
$-0.0939^b$&$-0.0192^b$\\
&$-0.1526^c$& & &$-0.1531^d$\\ \hline
$\Delta_{\textrm{sp},\mathrm{sp}}(3,n)/n$&$-0.0224^a$&
$-0.0252^a$&
$-0.0278^a$&$-0.0619^a$\\
&$-0.0117^b$&$-0.0111^b$&$-0.0106^b$&
$-0.0059^b$\\ \hline\hline
\end{largetabular}
\end{center}
$^a$ Values obtained by setting $\epsilon=1$ in Eqs.~(\ref{eq:Delper})
and (\ref{eq:Delspsp}).\\
$^b$ $O(\epsilon)$ results \cite{KD92a}, evaluated at $\epsilon=1$.\\
$^c$ MC results according to \cite{Kre97}.\\
$^d$ Exact
  value $-2\zeta(3)/(5\pi)$ according to \cite{Dan98} and
  \cite{DDG06}.
\end{table}
of $\Delta_\wp$ Eqs.~(\ref{eq:Delper}) and (\ref{eq:Delspsp}) predict
for $n=1,2,3,\infty$ upon setting $\epsilon=1$. For comparison, the
corresponding $O(\epsilon)$ estimates are also listed, along with a MC
estimate \cite{Kre97} and an exact $n=\infty$ result
\cite{Dan98,DDG06}. A known problem of the $O(\epsilon)$ results for
$\Delta_{\mathrm{per}}$ is the seemingly incorrect $n$-dependence of
the predicted $d=3$ values, whose deviations from the exact
${n=\infty}$ value \emph{increase} monotonically as $n$ grows (see
Fig.~12.8 of \cite{BDT00} and \cite{DK04}), although the MC estimate
for ${n=1}$ is very close to the exact ${n=\infty}$ value. The
$\epsilon^{3/2}$ term modifies the $n$~dependence, yielding an
estimate for $-\Delta_{\mathrm{per}}/n$ that increases with $n$.

In summary, we have shown the following: (i) the
$\epsilon$~expansions of quantities such as Casimir amplitudes are
ill-defined at $T_{c,\infty}$ when the BC gives a zero mode in Landau
theory.  (ii) The reformulation of field theory presented here yields
well-defined small-$\epsilon$ expansions for temperatures $T\geq
T_{c,\infty}$. In the cases of $\Delta_{\mathrm{per}}$ and
$\Delta_{\mathrm{sp},\mathrm{sp}}$, these expansions involve
fractional powers and logarithms of $\epsilon$.  Clearly, more work is
necessary to explore the potential of such expansions for reliable
extrapolations to $d=3$.

In typical experimental situations one expects to have either Robin BC
$\partial_n\bm{\phi}=\mathring{c}_j\bm{\phi}$ (which in the long-scale
limit normally should map on $(D,D)$ BC) or symmetry-breaking
$(+,\pm)$ BC (classical liquids). However, experimental situations
corresponding to near-critical enhancement of surface interactions on
both plates are conceivable. In that case a crossover from an initial
behavior characteristic of $(\mathrm{sp},\mathrm{sp})$ BC should
occur. Clearly, proper treatments of this crossover must take into
account the findings described above. Needless to say, that in MC
simulations periodic BC are \emph{the} preferred choice, and that
simulations dealing with $(\mathrm{sp},\mathrm{sp})$ BC were performed
as well \cite{PS98}.

\acknowledgments
We gratefully acknowledge discussions with D.~Dantchev, helpful
correspondence with M.~Krech, and partial support by Deutsche
Forschungsgemeinschaft via grant Die-378/5.

% \bibliographystyle{prsty}
% \bibliography{bank,casi,/home/hwd/DS06/rem}

\begin{thebibliography}{10}
\bibitem{rem:tc}
We use dimensional regularization. In a perturbative approach for small
  $\epsilon$, the critical value of $\mathring{\tau}$ remains at zero, so
  $\mathring{\tau}=0$ corresponds to $T=T_{c,\infty}$.
\bibitem{rem:cderi} In the case of special-special BC, we set both
  renormalized surface enhancement variables $c_j$, $j=1,2$, to their
  critical value $0$. When values $0<c_j<\infty$ are allowed,
  additional surface counterterms are required \cite{Die86a}.
  Subtracting from $f_L$ its Taylor expansion to the appropriate order
  in $\tau$ and $c_j$, one can fix these through normalization
  conditions for $f_{L,\mathrm{ren}}$ and its derivatives at
  $\tau=c_j=1$.
\end{thebibliography}

\end{document}